\documentclass[aps, floatfix, nofootinbib, preprintnumbers, showpacs, superscriptaddress]{revtex4}

\usepackage{amsfonts}
\usepackage{amssymb}
\usepackage{latexsym}
\usepackage{graphicx}
\usepackage{ifthen}
\usepackage{mathpple}

\newif\ifcolour
\colourtrue

\begin{document}

%%%%%%%%%%%%%%%%%%%%%%%%%%%%%%%%%%%%%%%%%%%%%%%%%%%%%%%%%%%%%%%%%%%%%%%%
% Begin title & abstract
%%%%%%%%%%%%%%%%%%%%%%%%%%%%%%%%%%%%%%%%%%%%%%%%%%%%%%%%%%%%%%%%%%%%%%%%
\title{Evolutions in 3D numerical relativity using fixed mesh refinement}

\author{Erik Schnetter}
\affiliation{Institut f\"ur Astronomie und Astrophysik,
             Universit\"at T\"ubingen,
             Auf der Morgenstelle,
             D-72076 T\"ubingen, Germany}
\altaffiliation{Max-Planck-Institut f\"ur Gravitationsphysik, 
             Albert-Einstein-Institut, D-14476 Golm, Germany}
\email{schnetter@aei.mpg.de}

\author{Scott H. Hawley}
\affiliation{Center for Relativity, 
             University of Texas at Austin, 
             Austin, TX 78712, USA}
\email{shawley@physics.utexas.edu}

\author{Ian Hawke}
\affiliation{Max-Planck-Institut f\"ur Gravitationsphysik, 
             Albert-Einstein-Institut, D-14476 Golm, Germany}
\email{hawke@aei.mpg.de}

\begin{abstract}
  We present results of 3D numerical simulations using a finite
  difference code featuring fixed mesh refinement (FMR), in which a
  subset of the computational domain is refined in space and time.  We
  apply this code to a series of test cases including a robust
  stability test, a nonlinear gauge wave and an excised Schwarzschild
  black hole in an evolving gauge.  We find that the mesh refinement
  results are comparable in accuracy, stability and convergence to
  unigrid simulations with the same effective resolution.  At the same
  time, the use of FMR reduces the computational resources needed to
  obtain a given accuracy.  Particular care must be taken at the
  interfaces between coarse and fine grids to avoid a loss of
  convergence at higher resolutions, and we introduce the use of
  ``buffer zones'' as one resolution of this issue.  We also introduce
  a new method for initial data generation, which enables higher-order
  interpolation in time even from the initial time slice.  This 
  FMR system, ``Carpet'', is a
  driver module in the freely available Cactus computational
  infrastructure, and is able to endow generic existing Cactus simulation
  modules (``thorns'') with FMR with little or no extra effort.
\end{abstract}


\preprint{AEI-2003-078}

\maketitle
%%%%%%%%%%%%%%%%%%%%%%%%%%%%%%%%%%%%%%%%%%%%%%%%%%%%%%%%%%%%%%%%%%%%%%%%
% End title & abstract
%%%%%%%%%%%%%%%%%%%%%%%%%%%%%%%%%%%%%%%%%%%%%%%%%%%%%%%%%%%%%%%%%%%%%%%%

%%%%%%%%%%%%%%%%%%%%%%%%%%%%%%%%%%%%%%%%%%%%%%%%%%%%%%%%%%%%%%%%%%%%%%%%
% Section: Introduction
%%%%%%%%%%%%%%%%%%%%%%%%%%%%%%%%%%%%%%%%%%%%%%%%%%%%%%%%%%%%%%%%%%%%%%%%
\section{Introduction}
Currently many researchers are involved in efforts to predict the
gravitational waveforms emitted by the inspiral and merger of compact
binaries.  The direct numerical simulation of these binary sources has
been regarded as the best, if not only, way to obtain information
about the waves emitted during the merger event itself.  Many
numerical simulation methods are currently in use, and among the most
popular of these is the use of finite difference approximations to
Einstein's equations on a uniform mesh or grid.

It is often the case in simulations of physical systems that the most
interesting phenomena may occur in only a subset of the computational
domain.  In the other regions of the domain it may be possible to use
a less accurate approximation, thereby reducing the computational
resources required, and still obtain results which are essentially
similar to those obtained if no such reduction is made.  In
particular, we may consider using a computational mesh which is {\em
  non-uniform} in space and time, using a finer mesh resolution in the
``interesting'' regions where we expect it to be necessary, and using
a coarser resolution in other areas.  This is what we mean by {\em
  mesh refinement} (MR).  In the case of large-scale simulations,
where memory and computing time are at a premium, the effective use of
MR can allow for simulations achieving effective (fine-grid)
resolutions which would be, on a practical level, impossible
otherwise.  This also implies that some systems requiring a large
dynamic range of resolution may only be possible to simulate using
mesh refinement.  One hopes that the use of fewer total grid points in
an efficient mesh refinement application, as opposed to a uniform,
single grid or {\em unigrid} application, may result in a saving of
computing time required to arrive at a solution. It is always the
case, however, that the reduction of memory requirements afforded by
mesh refinement, especially in 3D, could allow for highly accurate
simulations which simply may not fit in the memory of a given computer
in a unigrid setting.

In many simulations it may be desirable for the computational mesh to
{\em adaptively} refine in response to some criterion such as the
local truncation error.  Such {\em adaptive mesh refinement} (AMR)
systems provide, ideally, a very high degree of computational
efficiency for a given problem.  AMR schemes have been in use for
decades~\cite{BandO}, particularly in the fluid dynamics
community~\cite{BergerColella, FluidsAMR}.  Applications to to
astrophysical systems have been underway for quite some time
(e.g.,~\cite{Norman1, Norman2, Plewa, Khokhlov}).  In numerical
relativity, Choptuik~\cite{Matt1,Matt2} introduced many researchers to
AMR through his results using a 1D code for studying critical
phenomena.  Since the early 90's efforts have been underway in the
numerical relativity community to develop and employ AMR applications
in two and three dimensions, in studies of waves ~\cite{Dale,Hawley,
  Gerd, Joan, LeeWild, KimNew}, critical collapse ~\cite{Matt2, Steve,
  Pretorius, Choptuik03d}, the initial data
problem~\cite{Khokhlov,Diener, Lowe}, inhomogeneous cosmologies
~\cite{Hern}, Schwarzschild black holes ~\cite{Bruegmann96, Gerd,
  Breno}, characteristic methods ~\cite{FransLuis}, and a binary black
hole system~\cite{Bruegmann97}.  In several of these cases, the use of
mesh refinement provided not only a more efficient means of computing a
solution, but even allowed the authors to obtain solutions which would
not have been {\em possible} via standard unigrid methods, given
existing computing technology at the time.

One step in the development of an AMR code is the ability to handle
regions of different resolution which are known in advance --- {\em
fixed} mesh refinement (FMR).  Thus we view our present work as in
some sense a precursor to full 3D AMR simulations.  However, the FMR
code may be sufficient in its own right, as full AMR may not be
necessary for problems where the region of interest is well known
beforehand.

An important point should be made regarding the benefits of mesh
refinement for reducing storage requirements (and computational costs).
It is well known that pseudospectral collocation methods offer 
exponential convergence as the number of collocation points is increased
(for sufficiently smooth functions, which are present in many systems
of interest to numerical relativists),
and such methods are able to yield extremely high accuracies via
simulations which require storage afforded by a typical desktop
computer \cite{Kidder,Kidder2,Pfeiffer,Scheel,Scheel2}.  
Since FMR makes use of finite difference methods, we can obtain 
at best only polynomial convergence as we increase the resolution of
all grids.   Our choice of finite difference methods and FMR are 
based on a choice of physical systems of interest, and on
an observation.

One reason we are motivated to use mesh refinement is that we are
interested in systems which are ``non-smooth'' by the standards of
pseudospectral methods:  hydrodynamics, and gravitational collapse.
In hydrodynamics, the formation of discontinuities  --- shocks ---
from smooth initial data is a generic phenomena.
In gravitational collapse, features may appear on smaller
and smaller spatial scales, requiring a means of resolving these
features as they appear, either via a truly adaptive algorithm (in
which, e.g.,  the truncation error is used as a refinement criterion
\cite{BandO}) or a ``progressive'' mesh refinement system in which
nested grids are ``turned on'' at smaller radii and higher resolutions
as the simulation proceeds.

The observation mentioned above is that 
unigrid finite difference codes are generally
regarded as being significantly easier to develop than pseudospectral
codes.  Our interest, from a code-development standpoint, has been
to provide an infrastructure whereby existing unigrid codes can be
``endowed'' with mesh refinement in a way which is somewhat automatic,
such that the developer of the original unigrid code is spared the details
of implementing mesh refinement.  {\em If the introduction of mesh
refinement does not significantly alter the dynamics of the system},
then one should be able to obtain results comparable to a high
resolution unigrid run via an FMR run with appropriately placed
fine grids which share the resolution of the unigrid run.  In fact,
this is the hope of this paper, and the criterion by which we
evaluate our results obtained by FMR:  {\em ideally, the use of
mesh refinement should produce results of comparable quality to 
a corresponding
unigrid run}, in terms of stability, accuracy and convergence behaviour.
Thus our mesh refinement infrastructure could, in applicable cases,
provide a service to the community of researchers who commonly develop
unigrid finite difference codes, by providing these researchers a means
by which to achieve more accurate results than their current computer
allocations allow.  It is in this spirit that we are making the
FMR system, called Carpet (authored by Erik Schnetter, with 
refinements offered by several others), freely available as a driver
thorn of the open-source Cactus computational infrastructure
\cite{cactus-grid, cactus-tools, cactus-review, cactus-webpages,Goodale02a}.

%%%%%%%%%%%%%%%%%%%%%%%%%%%%%%%%%%%%%%%%%%%%%%%%%%%%%%%%%%%%%%%%%%%%%%%%
% Section: Our method
%%%%%%%%%%%%%%%%%%%%%%%%%%%%%%%%%%%%%%%%%%%%%%%%%%%%%%%%%%%%%%%%%%%%%%%%
\section{Overview of our FMR method}

\subsection{Cactus and mesh refinement}

Cactus is an application framework that provides some computational
infrastructure such as parallelisation and disk I/O, and lets users
specify the modules, called ``thorns'', which contain all the
``physics''.  The main difference between an application framework and
a library of subroutines is \emph{control inversion}, meaning that it
is Cactus calling the users' routines, while the main program is part
of Cactus.  The part of Cactus that controls the creation of initial
data and the time stepping is called the \emph{driver}, which is a
thorn that interacts with the Cactus scheduler in order to determine
which routines are applied to which grid at what time.  Control
inversion has the important advantage that, by replacing the driver,
one can change a Cactus application from unigrid to mesh refinement
without rewriting any of the users' thorns.

In practice, the ability to use the same code with both unigrid and
mesh refinement drivers places restrictions on the implementation of a
routine.  For most evolution thorns these
modifications are at most minor. Some thorns, particularly those
solving elliptic equations,
require more substantial alterations. In
most circumstances the restrictions imposed are only technical points,
and it should be simple for any new code to be implemented to work
with a unigrid or mesh refinement driver interchangeably.
Our experience is also that analysis tools for e.g.\ wave extraction
or apparent horizon finding continue to work almost without changes.

\subsection{Mesh refinement method}

\begin{figure}[htbp]
\includegraphics[scale=0.5]{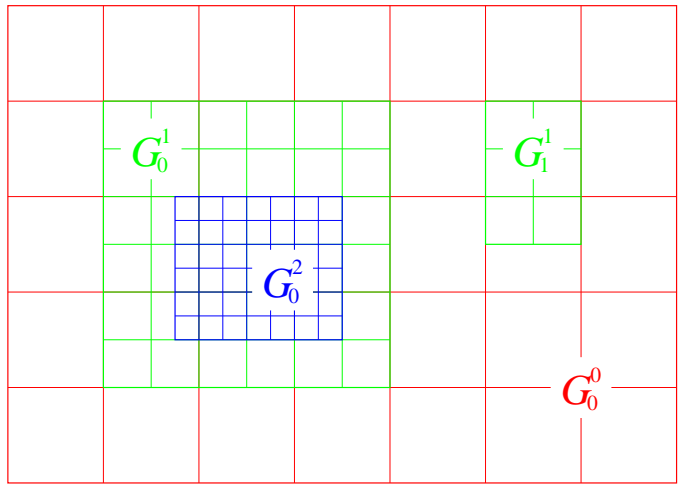}
\caption{
Base level $G^0_0$ and two refined levels $G^1_j$ and $G^2_j$, showing
the grid alignments and demonstrating proper nesting.}
\label{fig:Grids1}
\end{figure}

The mesh refinement driver that we use is called \emph{Carpet} and is
available together with the application framework Cactus.  It uses the
Berger--Oliger approach \cite{BandO}, where the computational domain
as well as all refined subdomains consist of a set of rectangular
grids.  In particular, we base our scheme on the so-called ``minimal
Berger--Oliger'' setup popularised by Choptuik \cite{MattMinBO}.
In this simplified version of Berger--Oliger, the grid points are 
located on a grid with Cartesian topology,
and the grid boundaries are aligned with the grid lines.
In our version, we also allow fine grid boundaries to occur 
in between coarse grid points, as shown in Figure~\ref{fig:Grids1}.
(Note that
this definition still allows, e.g., spherical coordinate systems.)
Furthermore, there is a constant refinement ratio between refinement
levels (described below).

We use the following notation. The grids are grouped into {\em
  refinement levels} (or simply ``levels'') $L^k$, each containing an
arbitrary number of grids ${G^k}_j$.  Each grid on refinement level
$k$ has the grid spacing (in one dimension) $\Delta x^k$.
The grid spacings are related by the relation
$\Delta x^k = \Delta x^{k-1} / N_{\text{refine}}$ with the integer
\emph{refinement factor} $N_{\text{refine}}$.
An example
is shown in Figure~\ref{fig:Grids1}.  In what follows we will assume
that $N_{\text{refine}}$ is always set to $2$.
The base level $L^0$ covers
the entire domain (typically with a single grid) using a coarse grid
spacing.  The base level need neither be rectangular nor connected.
The refined grids have to be \emph{properly nested}.  That is, any
grid ${G^k}_j$ must be completely contained within the set of grids
$L^{k-1}$ of the next coarser level, except possibly at the outer
boundaries.

Although Cactus does allow both vertex and cell centred grids,
current relativity thorns only use vertex centring.  Hence Carpet
currently only supports vertex centred refinement, i.e.\ 
coarse grid points coincide with fine grid points (where fine grid
points exist).

The times and places where refined grids are created and removed are
decided by some \emph{refinement criterion}.  The simplest criterion,
which is also indispensable for testing, is manually specifying the
locations of the refined grids at fixed locations in space at all
times. This is called \emph{fixed mesh refinement}.  A bit more
involved is keeping the same refinement hierarchy, but moving the
finer grids according to some knowledge about the simulated system,
tracking some feature such as a black hole or a neutron star.  This
might be called ``moving fixed mesh refinement''.  Clearly the most
desirable strategy is an automatic criterion that estimates the
truncation error, and places the refined grids only when and where
necessary.  This is what is commonly understood by \emph{adaptive mesh
  refinement}.  Carpet supports all of the above in principle, but we
will only use fixed mesh refinement in the following.
It should be noted that automatic grid
placement is a non-trivial problem (see,
e.g.,~\cite{Bell94,Berger91}).

\subsection{Time evolution scheme}
\label{sec:time-evolution}
\label{sec:boundary-prolongation}

\begin{figure}[htbp]
\includegraphics[scale=0.3]{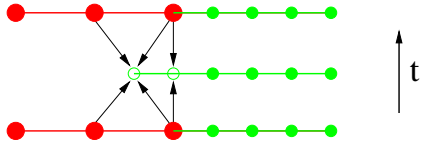}
\caption{Schematic for the prolongation scheme, in $1+1$ dimensions, for
a two-grid hierarchy.  The large filled
\ifcolour(red) \fi circles represent data
on the coarse grid, and smaller filled
\ifcolour(green) \fi circles represent data
on the fine grid.
The arrows indicate interpolation of coarse grid data in space and
time, necessary for the boundary conditions on the fine grid
(explained in section \ref{sec:boundary-prolongation}).}
\label{fig:Grids2}
\end{figure}

\begin{figure}[htbp]
\includegraphics[scale=0.3]{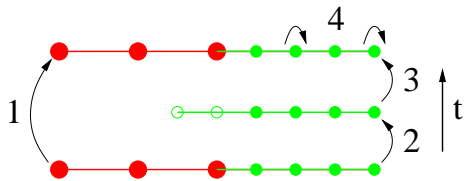}
\caption{Schematic for the time evolution scheme, in $1+1$ dimensions,
for a two-grid hierarchy.  The large filled
\ifcolour(red) \fi circles represent
data on the coarse grid, and smaller filled
\ifcolour(green) \fi circles represent
data on the fine grid.
The algorithm uses the following order.  1: Coarse grid time step, 2
and 3: fine grid time steps, 4: restriction from fine grid to coarse
grid.
Since the fine grid is always nested inside a coarse grid, there are
also coarse grid points (not shown) spanning the fine grid region (at
times when the coarse grid is defined) at the locations of ``every
other'' fine grid point; the data at these coarse grid points are
restricted (copied directly) from the fine grid data.}
\label{fig:Timestepping}
\end{figure}

The time evolution scheme follows that of the Berger and Oliger
\cite{BandO} AMR scheme, in which one evolves coarse grid data forward
in time before evolving any data on the finer grids.  These evolved
coarse grid data can then be used to provide (Dirichlet) boundary
conditions for the evolution of data on the finer grids via
\emph{prolongation}, i.e.\ interpolation in time and space.  This is
illustrated in Figure~\ref{fig:Grids2}.  For hyperbolic systems, where
a Courant-like criterion holds, a refinement by a factor of
$N_{\text{refine}}$ requires time step sizes that are smaller by a
factor $N_{\text{refine}}$, and hence $N_{\text{refine}}$ time steps
on level $k+1$ are necessary for each time step on level $k$.  At time
steps in which the coarse and fine grids are both defined, the fine
grid data are {\em restricted} onto the coarse grid (via a simple copy
operation) after it has been evolved forward in time.  If there are
more than two grid levels, then one proceeds recursively from coarsest
to finest, evolving data on the coarsest grid first, interpolating
this data in time and space along boundaries of finer grids, evolving
the finer grid data, and restricting evolved data from finer to
coarser grids whenever possible.  This is illustrated in
Figure~\ref{fig:Timestepping}.

For time evolution schemes that consist only of a single
iteration (or step), the fine grid boundary condition needs to be applied
only once.  Most higher-order time integrations schemes, such as
Runge-Kutta or iterative Crank-Nicholson, are actually multi-step
schemes and
correspondingly require the fine grid boundary condition to be applied
multiple times.  If this is not done in a consistent manner at each
iteration, then the coarse and the fine grid time evolution will not couple
correctly, and this can introduce a significant error.  We explain
this in more detail in Appendix~\ref{sec:bufferzones}.

\begin{figure}[htbp]
\includegraphics[width=0.3\textwidth]{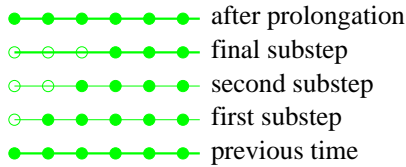}
\caption{Schematic for the ``buffering'' during time integration.
  Shown is the left edge of a refined region, which extends further to
  the right, which is integrated in time with a 3-step ICN method.  At
  the filled points (in the interior), time integration proceeds as
  usual.  The empty points (near the boundary) are left out, because
  no boundary condition is given during time integration.  A
  prolongation after the time integration fills the empty points
  again.  This whole scheme corresponds to either of the steps
  labelled 2 and 3 in Figure~\ref{fig:Timestepping}.}
\label{fig:bufferzones} 
\end{figure}

There are several ways to guarantee consistent boundary conditions on
fine grids. Our method involves not providing any boundary condition
to the individual integration substeps, but instead using a larger fine
grid boundary, as demonstrated in Figure~\ref{fig:bufferzones}.  That
is, each of the integration substeps is formally applied to a
progressively smaller domain, and the prolongation operation
re-enlarges the domain back to its original size.  Note that this
``buffering'' is done only for prolongation boundaries; outer
boundaries are handled in the conventional way.  Also, this is done
only for the substeps due to the time integration scheme, so that the
prolongation \emph{is} applied at fine grid times when there is no
corresponding coarse grid time. Note also that the use of buffer zones is
potentially more computationally efficient.

We emphasise that the use of these buffer zones is not always
necessary. To our knowledge the buffer zones are necessary only when
the system of equations contains second spatial derivatives,
\emph{and} a multi-step numerical method is used for time integration.
This issue arises for the BSSN system discussed below. We also give a
simple example using the scalar wave equation in
section~\ref{sec:waveeqn-periodic} and appendix~\ref{sec:bufferzones}.

\subsection{Inter-grid transport operators}

As described above, the interaction between the individual refinement
levels happens via prolongation and restriction.
For prolongation,
Carpet currently supports polynomial interpolation, up to 
quadratic interpolation in time, which requires keeping at least two
previous time levels of data.  It also supports
up to quintic interpolation in space,
which requires using at least three ghost zones.
We usually use cubic interpolation in space,
which requires only two ghost zones.
(Quadratic interpolation in space introduces an anisotropy
on a vertex centred grid.)
For restricting, Carpet currently uses sampling
(i.e., a simple copy operation). 
These transport
operators are not conservative.  Since our formulation of Einstein's
equation (see below) is not in a conservative form, any use of
conservative inter-grid operations offers no benefit.  However, the
transport operators can easily be changed.  (For more discussion of
the situations where conservative inter-grid operators are
useful or not, see~\cite{Quirk91}.)

\subsection{Initial data generation}
\label{section:InitialData}
Initial data generation and time evolution are controlled by the
driver Carpet.  Initial data are created recursively, starting on the
coarsest level $L^0$.  This happens as follows: On refinement level
$L^k$, the initial data routines are called.  This fills the grids on
this level.  Then the refinement criterion is evaluated (which might
be nothing more than a fixed mesh refinement specification).  If
necessary, grids on a finer level $L^{k+1}$ are created, and initial
data are created there, and on all finer levels recursively.  Then,
the data from level $L^{k+1}$ are restricted to level $L^k$ to ensure
consistency.

In many cases, the initial data specification is only valid for a
single time $t=0$, such as when using a time-symmetric approach, or
when solving an elliptic equation.  However, for the time
interpolation necessary during prolongation (see above), it may be
necessary to have data on several time levels.  One solution is to use
only lower order interpolation during the first few time steps.  We
decided instead, according to the Cactus philosophy, that the data
that are produced during the initial data creation should in principle
be indistinguishable from data produced by a time evolution.  Hence we
offer the option to evolve coarse grid data \emph{backwards in time}
in order to provide sufficient time levels for higher order
interpolation in time at fine grid boundaries.  This ensures that no
special case code is required for the first steps of the time evolution.

\begin{figure}[htbp]
\includegraphics[scale=0.6]{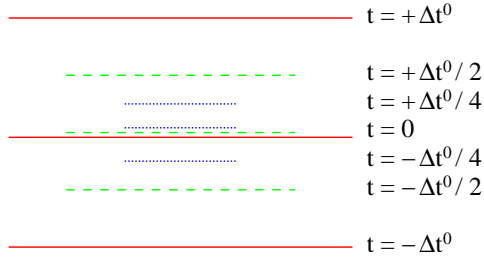}
\caption{Schematic for initial data scheme, in $1+1$ dimensions.  Our
use of quadratic interpolation in time requires three time levels of
coarse grid data in order to provide boundary data for evolution on
fine grids.  To achieve this from the beginning of the evolution
(without the use of a known continuum solution with which to
``pre-load'' these levels), we evolve our initial data (defined at
$t=0$) both forwards and \emph{backwards} one step in time.  In this
way, three time levels of coarse grid data are always available to
provide boundary data along the edges of fine grids.  The data at
various times are denoted by fractions of the time step $\Delta t^0$
on the base grid.  The coarsest grid is shown as by a solid
\ifcolour(red) \fi
line, a finer grid by a long-dashed
\ifcolour(green) \fi line, and a still finer
grid by a dotted \ifcolour(blue) \fi line.  (We
perform some additional backwards evolution as well, which we describe
in the main text.  The essence of the scheme, however, is given
here.)}
\label{fig:threelev_initdata1} 
\end{figure}

This initial data generation proceeds in two stages.  First the data
are evolved \emph{both forwards and backwards in time} one step,
leading to the ``hourglass'' structure illustrated
Figure~\ref{fig:threelev_initdata1}.  This evolution proceeds
recursively from coarsest to finest, so that all data necessary for
time interpolation are present.  Note that this would not be the case
if we evolved two steps backwards in time, as there would not
be enough data for the time interpolation for the restriction
operation between these two steps.

In the end we must provide initial data only at times
\emph{preceding} the initial time $t=0$; i.e., the hourglass
structure of Figure~\ref{fig:threelev_initdata1} is invalid as an
initial data specification in Cactus.  Therefore we perform in the
second stage of this scheme one additional step backwards in time on
each level, leading to initial data at the times $t_0$, $t_0-\Delta
t^k$, and $t_0-2\Delta t^k$ on each level $L^k$.

\section{Physical System}
The set of equations we solve are described in detail in
\cite{shiftpunctures}, and although we briefly review the material
here, we suggest interested readers refer to the
prior publication.  The evolution system is that of
Shibata-Nakamura~\cite{SN} and Baumgarte-Shapiro~\cite{BS}, the
so-called BSSN formulation.  The physical quantities present in a
typical ADM~\cite{York} evolution are the 3-metric $\gamma_{ij}$ and
the extrinsic curvature $K_{ij}$.  In the BSSN formulation, one
instead evolves a different set of variables: $K_{ij}$ is decomposed
into its trace $K$ and its trace-free part
\begin{equation}
  A_{ij} \equiv K_{ij} - {1\over 3} \gamma_{ij} K,
\end{equation}
and one applies a conformal transformation,
\begin{equation}
  \tilde \gamma_{ij} = e^{-4\phi} \gamma_{ij}.
\end{equation}
We choose $\phi$ such that the determinant of $\tilde\gamma_{ij}$,
denoted by $\tilde\gamma$, is $1$.

Thus our evolved quantities are related to the ADM and physical 
quantities by 
\begin{eqnarray}
  \phi &=& {1 \over 12} \log(\gamma) \\
  \tilde \gamma_{ij} &=& e^{-4\phi} \gamma_{ij} \\
  K &=& \gamma^{ij} K_{ij} \\
  \tilde A_{ij} &=& e^{-4\phi}(K_{ij} - {1 \over 3} \gamma_{ij} K)
\end{eqnarray}

One also creates a new evolved variable  $\tilde \Gamma^i$, defined
as
\begin{equation}
      \tilde \Gamma^i := \tilde \Gamma^i_{jk} \tilde \gamma^{jk}.
\end{equation}

The evolution equations for these variables are given by
\begin{eqnarray}
  \partial_t \phi &=& - {1 \over 6} \alpha K 
  + \beta^k \partial_k \phi 
  + {1 \over 6} \partial_k \beta^k
  \\
  \partial_t \tilde \gamma_{ij} &=& -2 \alpha \tilde A_{ij}
  + \beta^k \partial_k \tilde \gamma_{ij} 
  + \tilde \gamma_{ik}\partial_j \beta^k
  + \tilde \gamma_{jk} \partial_i \beta^k
  - {2 \over 3} \tilde \gamma_{ij}\partial_k \beta^k
  \\
  \partial_t K &=& 
%  -e^{-4\phi}( \tilde \gamma^{ij} \partial_i \partial_j \alpha - 
%  \tilde \Gamma^k \partial_k \alpha 
%  + 2 \tilde\gamma^{ij} \partial_i \phi \partial_j \alpha) 
  -D^i D_i \alpha
  + \alpha ( \tilde A_{ij} \tilde A^{ij} +
  {1 \over 3} K^2) 
  + \beta^i \partial_i K
  \\
  \partial_t \tilde A_{ij} &=&
   e^{-4 \phi} [-D_i D_j \alpha
  + \alpha R_{ij}]^{TF}
  + \alpha ( K \tilde A_{ij}
  - 2 \tilde A_{ik} \tilde A^k_{\ j}) \nonumber \\
  &&
  +  \tilde A_{kj} \partial_i \beta^k 
  +  \tilde A_{ki} \partial_j \beta^k  
  - {2 \over 3}  \tilde A_{ij} \partial_k \beta^k  \\
  \partial_t \Gamma^i &=&
  -2 \partial_j \alpha  \tilde A^{ij}
  + 2 \alpha (\tilde \Gamma^i_{jk} \tilde A^{kj} -
  {2 \over 3} \tilde \gamma^{ij} \partial_j K  +
  6 \tilde A^{ij} \partial_j \phi ) \nonumber \\
  &&
  - \partial_j (\beta^k \partial_k \tilde \gamma^{ij}
  - \tilde \gamma^{kj} \partial_k \beta^i
  - \tilde \gamma^{ki} \partial_k \beta^j 
  + {2 \over 3} \tilde \gamma^{ij} \partial_k \beta^k ),
\end{eqnarray}
where $D_i$ is the covariant derivative corresponding to the 3-metric
$\gamma_{ij}$, $R_{ij}$ is the three dimensional Ricci tensor, and the
``$TF$'' superscript denotes the trace-free part of the enclosed
expression.

The gauge conditions are given as follows.  The lapse $\alpha$ is
chosen as one of the Bona-Mass\'o \cite{Bona-Masso} family of
slicings,
\begin{equation}
  \partial_t \alpha = -\alpha^2 f(\alpha) (K-K_0) 
\end{equation}
where $K_0 \equiv K(t=0)$, and $f(\alpha)$ is chosen to give us either
harmonic slicing ($f(\alpha) = 1$) or a ``$1+\log$'' slicing
($f(\alpha) = 2 / \alpha$).  In this paper the shift will be held
constant at the analytic value in all cases.

We use the code described in~\cite{Alcubierre00a,Alcubierre01a} and
refer there for details of the finite differencing scheme used.

For
certain problems a small amount of artificial dissipation is useful.
We use a dissipation of the form
\begin{eqnarray}
   \partial_t\, u & = & - \epsilon\,h^4\,
      ( \partial_{xxxx} + \partial_{yyyy} + \partial_{zzzz} )\, u
\end{eqnarray}
with the grid spacing $h$,
which is described in \cite{kreiss-oliger}.  Although this dissipation
operator is typically employed only in systems with first order
derivatives in time and space, we find that its use in the BSSN system
(which has first order derivatives in time, and second order
derivatives in space) is effective at reducing high-frequency
oscillations (i.e., noise) in the simulations, but has little effect
on the overall convergence behaviour.

%%%%%%%%%%%%%%%%%%%%%%%%%%%%%%%%%%%%%%%%%%%%%%%%%%%%%%%%%%%%%%%%%%%%%%%%
% Section: Tests/Results
%%%%%%%%%%%%%%%%%%%%%%%%%%%%%%%%%%%%%%%%%%%%%%%%%%%%%%%%%%%%%%%%%%%%%%%%
\section{Tests}

For simplicity we will present tests using only two levels containing
one grid each, which we will refer to as the ``coarse grid'' and the
``fine grid''.  The fine grid is a box contained in the larger coarse
grid box, with the fine grid having a mesh spacing (in space and time)
of half that of the coarse grid. The only limitation on the number of
grids is the available computational resources, and we have
successfully performed tests with up to 24 levels of refinement.

One of the principal criteria we use to evaluate the effectiveness of
the FMR scheme is the requirement of second-order convergence in the
limit as the mesh spacing goes to zero.  Thus we run a given
simulation many times at different resolutions.  In our examples we
compare against the exact solution and check the convergence of the
solution error.
We show such tests only for the data on the coarsest grid, because the
restriction operator ensures that the coarse grid and fine grid data
are identical.

\subsection{Wave Equation: Periodic Boundaries}
\label{sec:waveeqn-periodic}
\begin{figure}
\includegraphics[width=0.45\textwidth]{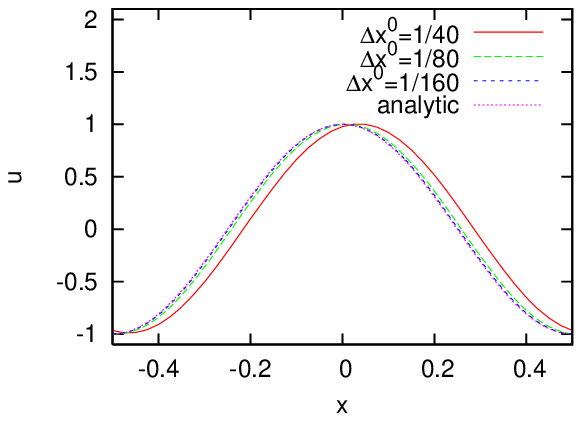}
\includegraphics[width=0.45\textwidth]{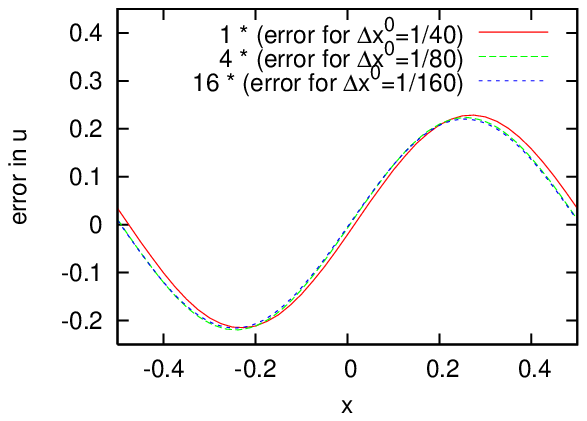}
\caption{
Result at $50$ crossing times
in the evolution of the wave equation
with periodic
boundaries, for two-level runs at three different resolutions ($1/40$,
$1/80$, and $1/160$).  Here
we show only data for the coarse grid;
the finer grid covers the box $x_i \in [-0.25,+0.25]$.
The left panel shows slices along the $x$-axis of the (3D) function $u$
found numerically, as well as the analytic (i.e.\ continuum) solution. (In
this graph the highest resolution data and the analytic solution appear
to lie on top of
each other.)  One can see a resolution-dependent phase shift in $u$.
The right panel shows the solution error, defined as the difference
between the $u$ obtained numerically and the continuum solution.  These
error graphs have been scaled to demonstrate the second order convergence of
the numerical results.
}
\label{fig:waveeqn}
\end{figure}

We tested our code with a simple wave equation in flat space in
Cartesian coordinates using several different kinds of initial data and
boundary conditions.  The first such test was that of sinusoidal plane
waves in a 3D box with periodic boundary conditions.  From a
code-development standpoint, we simply took an existing set of
subroutines for solving the wave equation in parallel and ran using
the FMR driver instead of the usual unigrid driver.  The formulation
of the wave equation we used was a single equation with second order
derivatives in both time and space, i.e.\ 
\begin{equation}
   \partial_{tt} u = \partial^i \partial_i u,
\label{eqn:sw-secondorderts}
\end{equation}
which we solved using a leapfrog-like scheme.  Second order
convergence was found, as shown in Figure~\ref{fig:waveeqn}.

\begin{figure}
\includegraphics[width=0.45\textwidth]{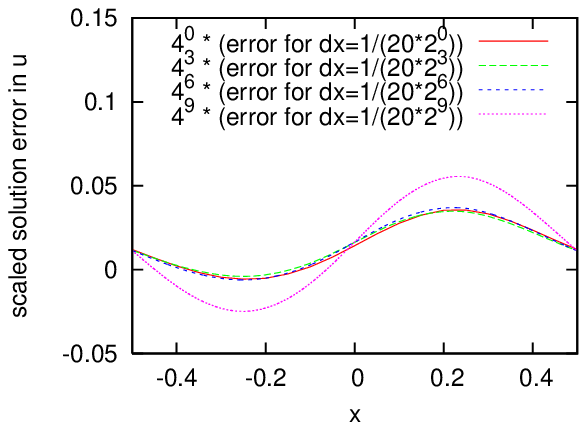}
\includegraphics[width=0.45\textwidth]{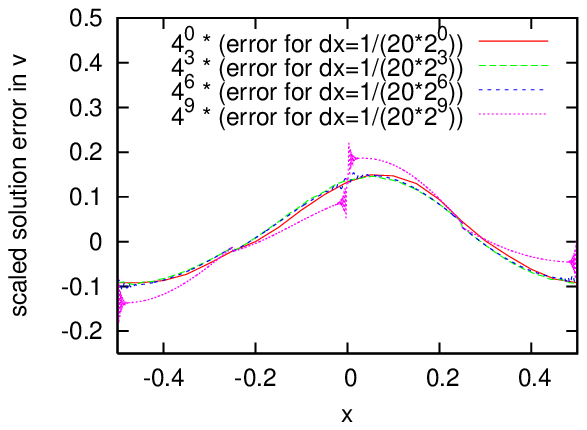}
\caption{
Scaled solution errors for the 1-D scalar wave equation at $0.75$ crossing
times, discretised \emph{without} buffer zones.  The left panel shows
the error in the solution $u$, the right panel in its time derivative
$v$.  At low resolutions, i.e.\ from $1/20$ up to about $1/1280$, the
scheme seems to be second-order convergent.  However, at higher
resolutions it becomes clear that this is not the case.  At lower
resolutions, the refinement boundaries are visible as small
discontinuities in the error of $v$ at $x=\pm0.25$.  At
higher resolutions, the discontinuity develops an oscillating tail and
propagates through the simulation domain.  It is instructive to see
that 8 convergence test levels were necessary to see this behaviour
numerically. Compare this against Figure~\ref{fig:waveeqn-good}.}
\label{fig:waveeqn-bad}
\end{figure}

\begin{figure}
\includegraphics[width=0.45\textwidth]{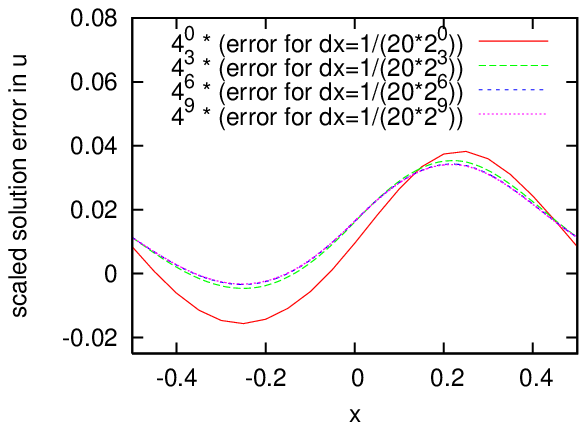}
\includegraphics[width=0.45\textwidth]{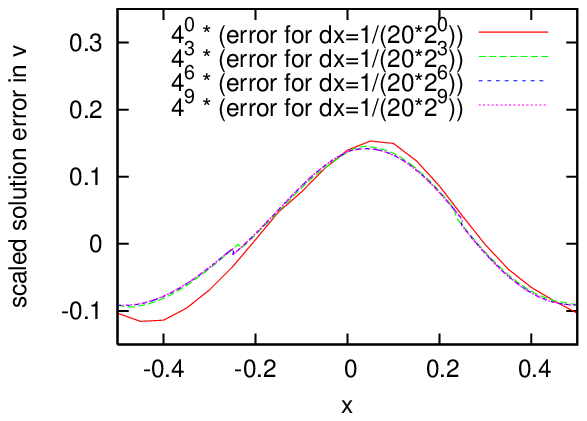}
\caption{
Scaled solution errors for the 1-D scalar wave equation at $0.75$ crossing
times, discretised \emph{with} buffer zones.  The left panel shows
the error in the solution $u$, the right panel in its time derivative
$v$.
The refinement boundaries are visible as small
discontinuities in the error of $v$ at $x=\pm0.25$.
The results are second-order convergent
up to extremely high resolutions.
It can also be seen that the resolution $1/20$
is clearly not yet in the convergence regime.}
\label{fig:waveeqn-good}
\end{figure}

An alternative formulation of the wave equation uses second order
derivatives in space but only first order in time. We write it in the form
\begin{eqnarray}
\label{eqn:sw-secondorder}
   \partial_t u & = & v \\
   \partial_t v & = & \partial^i \partial_i u
\end{eqnarray}
This formulation is comparable to the ADM (and BSSN)
formulations of the Einstein equations.
When this formulation of the scalar wave equation
is evolved using ICN integration without buffer
zones, the result is only first-order convergent, as shown using a one
dimensional example in Figure~\ref{fig:waveeqn-bad}. The same
formulation evolved with buffer zones converges even at extremely high
resolution as shown in Figure~\ref{fig:waveeqn-good}.

Having demonstrated the need for careful handling of refinement
boundaries, and having introduced buffer zones
as an effective approach to that, we use
buffer zones in all the remaining tests discussed in this paper.

\subsection{Wave Equation: Gaussian pulse}

We consider a Gaussian pulse that crosses a mesh refinement boundary,
travelling from the fine into the coarse region.  This is supposed to
mimic the case of gravitational waves propagating from fine, inner
grids and radiating out into coarser grids.

\begin{figure}
\includegraphics[width=0.45\textwidth]{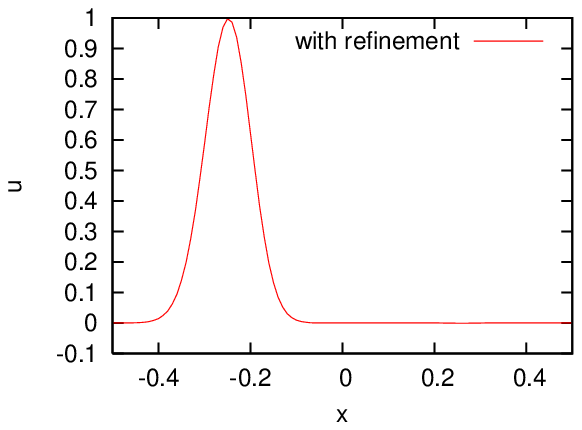}
\includegraphics[width=0.45\textwidth]{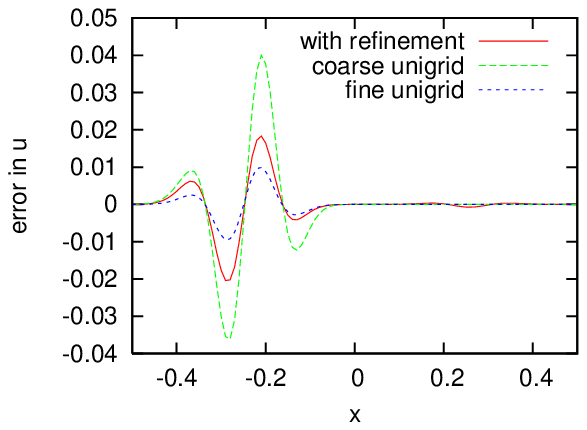}
\caption{
The left panel shows a Gaussian pulse that has travelled through a
refinement boundary located at $x=0$.  The region $x>0$ is refined.
The right panel shows the difference to the analytic solution, and
shows two unigrid runs for comparison.  The resolutions of the unigrid
runs correspond to the coarse and the fine regions in the mesh
refinement run.  As expected, the error in the run with refinement is
in between the errors of the two unigrid runs.  The reflected part of
the pulse is very small; it has an amplitude of only $10^{-3}$
and is thus much smaller
than the discretisation error in the pulse itself.}
\label{fig:pl1d}
\end{figure}

We use an effectively one dimensional domain (planar symmetry in 3D)
with $x \in [-0.5, +0.5]$, and a coarse grid resolution of $\Delta
x^0=1/100$.  The region $x>0$ is refined by a factor of $2$.  The
Gaussian pulse starts in the refined region and travels to the left.
Figure~\ref{fig:pl1d} shows the pulse after it has crossed the
interface, and compares the result to two unigrid simulations.  The
errors that are introduced at the refinement boundary are very small
and converge to second order. In particular, at this resolution about
$10^{-3}$ of the original pulse is reflected by the refinement
boundary. 

\subsection{Wave Equation: $1/r$, with excision}
Another test we performed was the solution of the wave
equation~(\ref{eqn:sw-secondorderts}) using initial data which scaled
inversely with $r$, the distance from the origin.  This was conceived
as a useful step before moving on to black holes
\cite{Bernd_one_over_r_comment}, as the ``puncture data''
\cite{Brandt97b} we use for the black holes has elements which scale
as $1/r$ in the vicinity of the puncture. Other systems with isolated
central masses may also be expected to have elements which scale in a
similar fashion.  The $1/r$ data are a static solution of the wave
equation in 3D, and are compatible with the standard Sommerfeld
outgoing boundary condition.  To handle the singularity at the centre,
we ``excise'' the centre of the computational domain by choosing some
inner boundary at a finite {\em excision radius} from the centre and
filling the interior region with prescribed data.  Thus in this test
we also have a test of the use of FMR in the presence of excision,
which is commonly used for the interiors of black holes in analogous
simulations.  We perform no evolution within this excised region.

\begin{figure}
  \centerline{
    \includegraphics[width=0.4\textwidth]{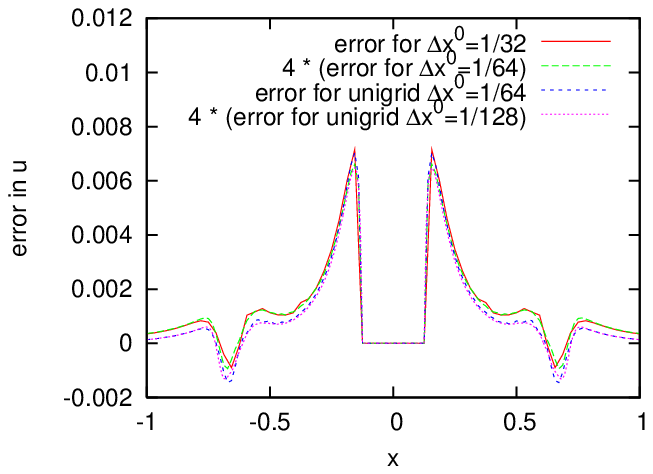}
    \hspace{0.25cm}
    \includegraphics[width=0.4\textwidth]{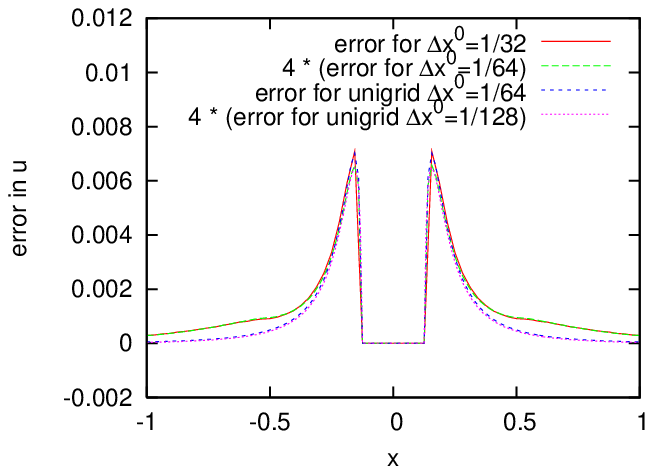} }
\caption{
Solution error for evolutions of $1/r$ data at times $t=0.625$ 
(left frame) and $t=5.0$ (right frame), using
radiation (Sommerfeld)
boundary conditions.  Here we show only the base
grid; the refined grid covers the domain $x_i \in [-0.5, +0.5]$.
The region $|x_i| \le 0.125$ is excised.
We see in the left frame that an initial transient moving outward 
from the refined region is still resolved in a second-order-convergent
manner after it passes through the refinement boundary.
At later times, the error settles down to the profile shown in the 
right frame. 
As expected, the FMR results compare favourably with the unigrid 
results in the region covered by the refined grid. 
(Note that, outside the refined region, the errors shown for the FMR results 
have effectively been multiplied by a factor of 4 relative to the comparable 
unigrid results, giving the impression of greater disagreement than is really present.)
}
\label{fig:one_over_r}
\end{figure}

We use a full 3D grid with $x_i \in [-1, +1]$ and coarse grid
resolutions of $\Delta x^0 = 1/32$ and  $\Delta x^0 = 1/64$.
The region $x_i \in [-0.5, +0.5]$ is
refined by a factor 2, and the region $|x_i| \le 0.125$ is excised.
Graphs of the error at two different times in
the evolution are shown in Figure~\ref{fig:one_over_r},
which also shows corresponding unigrid runs for comparison.
We see that
the solution is fully convergent, and similar to the corresponding
unigrid results in the region of refinement.
It is interesting to note that even outside this region,
the FMR and unigrid results are very similar for
the ``transient'' shown in the left frame, however the late time
(``stationary'') behaviour shown in the right frame reveals a 
notable difference between the FMR and unigrid results outside 
the refined grid.

Having demonstrated the existence of convergent solutions of the
wave equation for oblique angles of incidence to refinement boundaries
(in the $1/r$ case),
and convergent solutions in which the amplitude of the reflected
wave is significantly (roughly three orders of magnitude) less than
the amplitude of the transmitted wave, we now move on to full
solutions of Einstein's equations.  For more detailed calculations
of reflection and transmission effects at refinement boundaries
see~\cite{Gerd,Joan}.

\subsection{Robust Stability}
\label{sec:robust-stability}

We have applied a robust stability test \cite{robust-stability,
  robust-stability-2, Mexico1} to the BSSN formulation.  This is a
test that is meant to supplement and extend an analytic stability
analysis, especially in cases where such an analysis is difficult or
impossible (because the equations contain too many terms).  A
numerical test has the advantage that it tests the complete
combination of evolution equations, gauge conditions, boundary
conditions, as well as the discretisation and the implementation, and
(in our case) the mesh refinement scheme.  Thus, while a numerical
test is not as reliable as an analytically obtained statement, it is
able to cover more general cases.

The robust stability test proceeds in three stages of increasing
difficulty:
\begin{description}
\item[Stage I:] Minkowski (or some other simple)
  initial data with small random perturbations.  The simulation domain
  is a three dimensional box with periodic boundary conditions.
  The perturbations
  should be linear, so we chose a maximum amplitude of $10^{-10}$.
  The periodicity means that there is effectively no boundary, so that
  this stage is independent of the boundary condition.  A code is
  deemed to be robustly stable if it shows at most polynomial growth
  and if the growth rate is independent of the grid resolution.  This is
  different from other definitions of stability, where exponential
  growth is often deemed to be stable if the rate of exponential
  growth is independent of the grid resolution.
\item[Stage II:] The same as stage~I, except that the boundaries in
  the $x$ direction are now Dirichlet boundaries.  In addition to the
  noise in the initial data, noise is also applied to these Dirichlet
  boundaries.  This tests the consistency of the formulation with the
  boundary conditions, but without the complications of edges and
  corners.
\item[Stage III:] The same as stage~II, except that there are now
  Dirichlet boundary conditions in all directions.
  This tests whether edges and
  corners are handled correctly by the combination of the formulation
  and boundary conditions.
\end{description}

In accordance with \cite{Mexico1} we chose an effectively one
dimensional, planar symmetric domain that extends almost only in the
$x$-direction with $x \in [-0.5, +0.5]$.
The domain has only 3 grid points in the $y$ and $z$ directions.
Thus we have to omit stage~III of the test here.
We used a resolution of $1/50$, and refined the centre of the domain.
Although the domain was thus essentially one dimensional, the
simulation was performed with the full three dimensional code.

\begin{figure}
\includegraphics[width=0.45\textwidth]{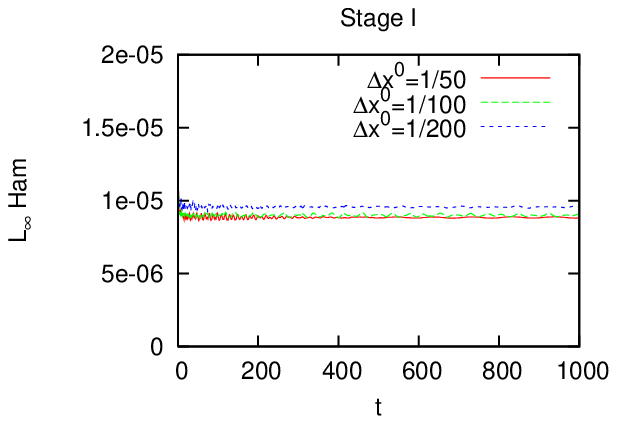}
\includegraphics[width=0.45\textwidth]{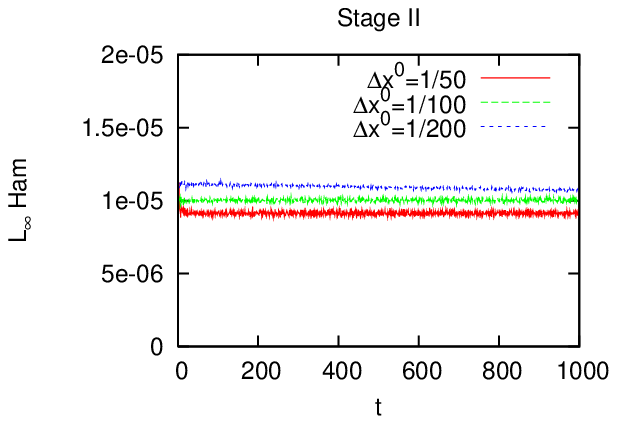}
\caption{
$L_\infty$-norm of the Hamiltonian constraint vs.\ time for the robust
stability test.  The left hand side graph shows stage~I where all
boundaries are periodic.  The right hand side graph shows stage~II,
which has noisy Dirichlet boundaries in the $x$ direction.  The domain
is essentially one dimensional with $x \in [-0.5, +0.5]$, and
periodic in the $y$ and $z$ directions.  The refined grid covers
the region $x \in [-0.25, +0.25]$.  The constraint violation in
corresponding unigrid runs (not shown) has the same magnitude.}
\label{fig:rs-test}
\end{figure}

Figure~\ref{fig:rs-test} shows
the $L_{\infty}$-norm of the Hamiltonian constraint versus time for
$1000$ crossing times for the stages~I and~II
for three different resolutions.
% and compares the refinement runs to corresponding unigrid runs.
All test were run
without artificial dissipation.  The results show that our
implementation is robustly stable.

\subsection{Gauge Wave}
As another test of the BSSN system we implemented the gauge wave
of~\cite{Mexico1}.  From the typical Minkowski coordinates $\{\hat t,
\hat x, \hat y, \hat z\}$, one defines new coordinates $\{t, x, y,
z\}$ via the coordinate transformation
\begin{eqnarray}
\hat t &=& t - {Ad\over 4\pi} \cos\left( {2\pi(x-t)\over d} \right) \\
\hat x &=& x + {Ad\over 4\pi} \cos\left( {2\pi(x-t)\over d} \right) \\
\hat y &=& y \\
\hat z &=& z,
\end{eqnarray}
where $d$ is the size of the simulation domain.
In these new variables, the 4-metric is
\begin{equation}
   ds^2 = - H dt^2 + H dx^2 + dy^2 dz^2, \label{GaugeWaveMetric}
\end{equation}
where 
\begin{equation}
   H = H(x-t) = 1 - A \sin\left( {2\pi (x-t) \over d} \right).
\end{equation}

This test provides us with an exact solution to which we can compare
our numerical results.  In addition to the exact values $\alpha =
\sqrt{H}$ and $g_{xx} = H$, we will compare the extrinsic curvature
$K_{ij}$, for which the only nonzero component is
\begin{equation}
 K_{xx} = - {\pi A \over d \sqrt{H}} \cos\left( {2\pi(x-t)\over d} \right).
\end{equation}
Since $\beta^i=0$ in the analytic solution we do not evolve the shift
but keep it set to zero at all times.

\begin{figure}
\includegraphics[width=0.45\textwidth]{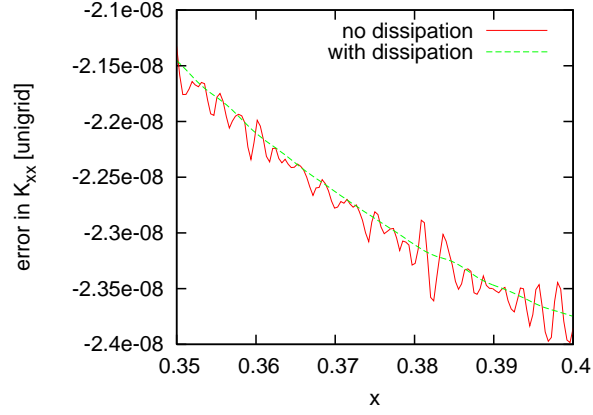}
\caption{
  Error in the extrinsic curvature component $K_{xx}$
  along a small part of the $x$ axis for a {\em unigrid}
  simulation after $0.25$ crossing times.  The resolution, $\Delta
  x=1/2560$, is rather high.  The noise (that was not present in the
  initial data) can be significantly reduced by dissipation.  It is
  surprising to see that a unigrid simulation with the plain BSSN
  formulation shows this behaviour; this might point to an instability
  in the system of equations.}
\label{fig:gw-kxx-uni}
\end{figure}

For this simulation we find it useful to add dissipation to the
evolution equations to suppress high frequency noise at very high
resolutions.  The reason is evident from unigrid simulations at high
resolutions, as demonstrated in Figure~\ref{fig:gw-kxx-uni}.

\begin{figure}
\includegraphics[width=0.45\textwidth]{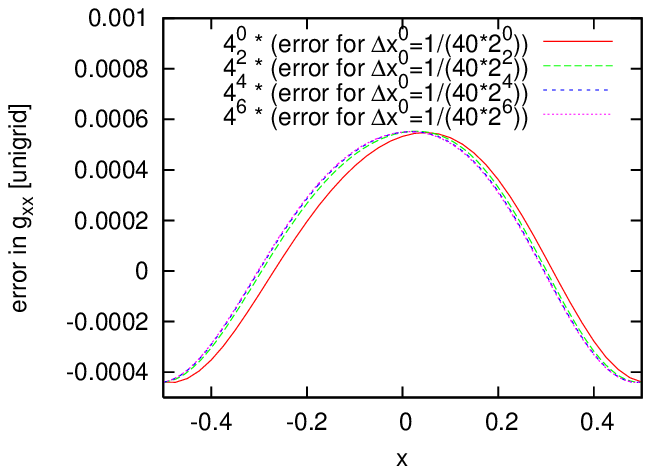}
\includegraphics[width=0.45\textwidth]{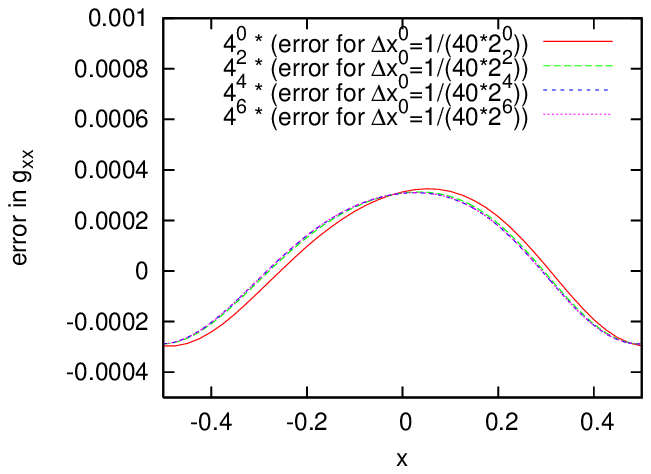}
\caption{
Errors for a uniform grid (left) and with mesh refinement (right)
after $5$ crossing
times for the three-metric component $g_{xx}$ in the gauge wave test
case.  The errors have been scaled with the resolution according to
second order convergence.}
\label{fig:gw-gxx}
\end{figure}

The simulation domain was set up in the same way as in \cite{Mexico1},
which is also the same as is used in Section~\ref{sec:robust-stability}.
That is, the simulation domain extended almost only in the $x$
direction with $x \in [-0.5, +0.5]$.  The domain has only 3 grid points in
the $y$ and $z$ directions.  All boundaries are periodic.
Although the domain is thus essentially one dimensional, the
simulation was performed with the full three dimensional code.
Figure~\ref{fig:gw-gxx} shows the results after $5$ crossing times for
the metric component $g_{xx}$, comparing refinement and unigrid runs.
We see perfect second order convergence.

\subsection{Schwarzschild Black Hole with Excision}
Next, we evolved a static Schwarzschild black hole in Kerr-Schild
coordinates which are manifestly static. The line element is
\begin{equation}
  \label{eq:EF}
  ds^2 = -\left(1-\frac{2M}{r}\right)\, dt^2
         + \left(\frac{4M}{r}\right)\, dt\, dr
         + \left(1+\frac{2M}{r}\right)\, dr^2 + r^2\, d\Omega^2
\end{equation}
where $M$ is the mass of the black hole. We choose to use excision to
handle the singularity. We use the excision method together with one
of the tests described in~\cite{Alcubierre00a}. The shift is held
static at the analytic solution and the lapse is evolved using $1+\log$
slicing. For simplicity we do not search for an apparent horizon but
merely excise those points within a cube with corners at $1M$. This
small excision region removes the divergences due to the singularity
while retaining some of the steep gradients. However, as noted by
\cite{excision-boundaries1,excision-boundaries2},
this does not guarantee that the future light cone
at the excision boundary is contained within the excised region.

We evolve only one octant of the grid to take advantage of the
symmetries present. We use a coarse grid spacing of $\Delta x^0 = 0.4M$
and $\Delta t^0 = 0.1M$ with $29^3$ points, giving an outer boundary at
$10M$ (ignoring two symmetry points and two outer boundary
points). The fine grid also contains $29^3$ points. The points are not
staggered about the origin, so there is always a grid point at $r=0$
which must be excised.

We compare the runs using mesh refinement with simulations using the
unigrid code as described in~\cite{Alcubierre00a,Alcubierre01a}.  The
coarse unigrid test is identical to the coarse grid in the simulation
using refinement. That is, $29^3$ points are used with a grid spacing
of $\Delta x = 0.4M$ giving an outer boundary at $10M$.  For the
medium unigrid run the resolution is doubled whilst the outer boundary
location is held fixed, giving the same effective resolution near the
excision region as the simulation using refinement. That is, $55^3$
points are used with a grid spacing of $\Delta x = 0.2M$. To compare
with the high resolution run using refinement we also perform a
unigrid run with the same effective resolution. That is, $105^3$
points are used with a grid spacing of $\Delta x = 0.1M$.

\begin{figure}[htbp]
    \includegraphics{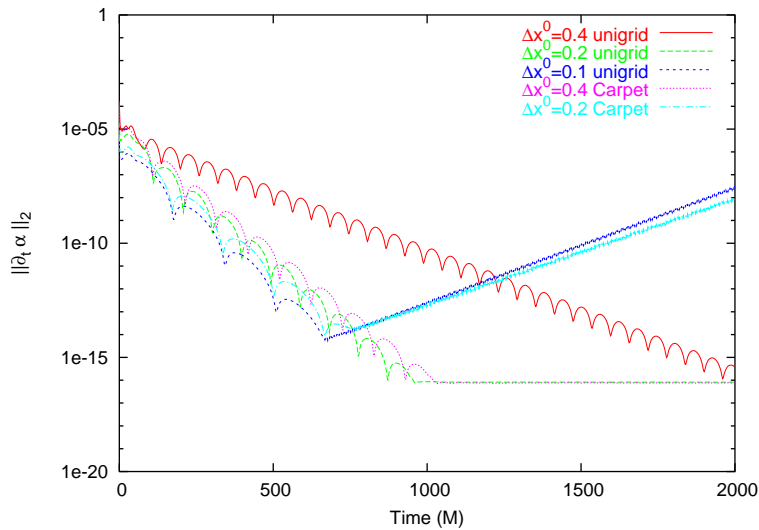}
    \caption{The r.m.s.\ of the change in the lapse with time. As
      in~\cite{Alcubierre00a} we see an exponentially damped
      oscillation as the system settles down. However, at sufficiently
      high resolution an instability sets in. This appears to come
      from floating point round-off error at the initial time and is
      clearly not caused by mesh refinement.}
    \label{fig:Excision1}
\end{figure}
Results showing the norm of the change of the lapse in time are shown in
Figure~\ref{fig:Excision1}. Firstly we note that the use of refinement
combined with an excision boundary has no qualitative effect on the
simulation. As in~\cite{Alcubierre00a} the change in the lapse shows
an exponentially damped oscillation in the absence of instabilities. 

However, in this case we do see an exponentially growing instability
which sets in only at the highest resolutions. By tracing back the
magnitude of this mode we see that it appears to come from floating
point round-off error at the initial time. It appears in runs both
with and without mesh refinement and the growth rate is the same in
both cases. The origin of this instability is unclear, especially as
very similar simulations with the same resolutions were shown to be
stable in~\cite{Alcubierre00a}. However, the important point for this
paper is that the stability of the simulations are independent of the
mesh refinement.

\begin{figure}[htbp]
    \includegraphics[width=0.6\textwidth]{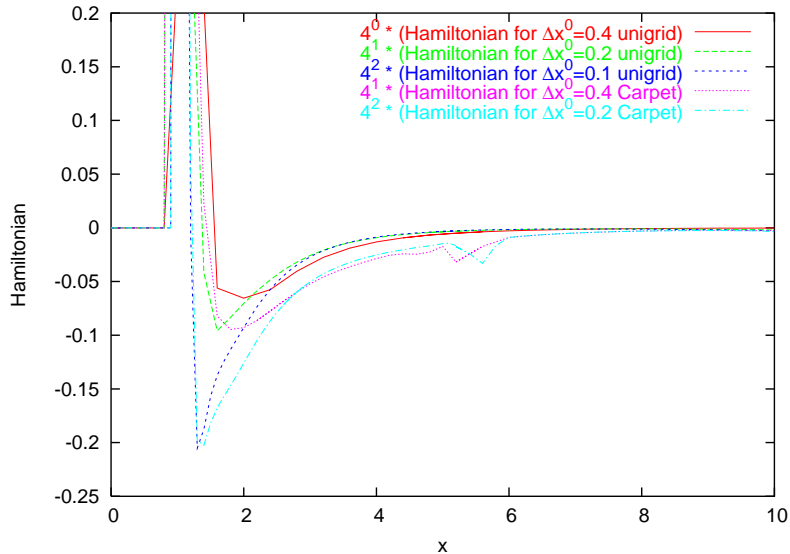}
    \caption{
      Second order convergence of the Hamiltonian constraint for the
      Schwarzschild black hole at $t=500M$. For the results using mesh
      refinement we plot the results using the finest possible level,
      causing a ``jump'' at the interface between the fine and coarse
      grid. We see that the mesh refinement simulations converge to second
      order except at the excision boundary. The simulation with
      refinement has comparable accuracy as the unigrid run with the
      same effective resolution near the excision boundary, but is
      less accurate away from this region.  }
    \label{fig:Excision2}
\end{figure}

The convergence of the Hamiltonian constraint at $t=500M$ is shown in
figure~\ref{fig:Excision2}, which should be compared with Figure 3
in~\cite{Alcubierre00a}. Second order convergence for the unigrid and
mesh refinement simulations are clear away from the excision boundary,
whilst at the excision boundary the convergence is not so obvious due
to the low resolution. The error in the mesh refinement runs is
comparable to the unigrid runs with the same effective resolution,
although only near the excision boundary are the full benefits seen.
Given the computational resources required, illustrated in
Table~\ref{tab:efficiency}, the benefits of mesh refinement are clear.

\begin{table}[htbp]
  \begin{center}
    \begin{tabular}[c]{|c|c|c|c|c|}
      \hline
      Effective & Grid size & Refinement &
      Memory & Real (user) time \\
      resolution (M) & & levels & use (MB) & (s)  \\ \hline
      0.4 & 25  & 1 & 58   & 18.0   (17.6)   \\
      0.2 & 25  & 2 & 114  & 85.5   (84.7)   \\
      0.2 & 50  & 1 & 303  & 222.1  (219.4)  \\
      0.1 & 50  & 2 & 695  & 957.6  (950.8)  \\
      0.1 & 100 & 1 & 2024 & 3337.9 (3309.9) \\ \hline
    \end{tabular}
    \caption{Computational resources required to evolve the single
      black hole with excision to $t=2M$ with various grid sizes. The
      results using refinement give comparable results to unigrid
      results with the same effective resolution whilst using
      approximately 30\% of the resources. These results used a 2.8GHz
      Intel Xeon processor with the Intel version 7 compilers.}
    \label{tab:efficiency}
  \end{center}
\end{table}

%%%%%%%%%%%%%%%%%%%%%%%%%%%%%%%%%%%%%%%%%%%%%%%%%%%%%%%%%%%%%%%%%%%%%%%%
% Section: Conclusions
%%%%%%%%%%%%%%%%%%%%%%%%%%%%%%%%%%%%%%%%%%%%%%%%%%%%%%%%%%%%%%%%%%%%%%%%
\section{Conclusions}

We have shown comparisons of results obtained via fixed mesh
refinement (FMR) techniques and via standard unigrid methods.
Ideally, the appropriate use of FMR should produce results very
similar in accuracy to the corresponding unigrid results, while
retaining the stability and convergence properties of the unigrid
simulations.  ``Corresponding'' in this case implies that the unigrid
simulation has the same resolution as the finest grid of the FMR
simulation. This ideal is only expected to hold for simulations where
most of the ``interesting behaviour'' occurs in a limited region which is
covered by refined grids, such as in simulations of compact objects.

We find that particular care must be taken at boundaries between
coarse and fine grids, in order to retain stability and convergence.
In cases where the system contains second spatial derivatives and
a multistep time integration method is used, we introduce
buffer zones to guarantee the convergence of the mesh refinement
implementation.  The need for these buffer zones can be clearly
seen in very fine resolution simulations of the wave equation when
written in ``mixed'' form (with second spatial derivatives and first
derivatives in time).  The same effects are seen in simulations of
gauge waves
even at intermediate resolutions
when using the BSSN formulation in a similar mixed form.

In order to use parabolic interpolation in time, even from the initial
time, we designed a scheme by which data is initially evolved forwards
and backwards in time.  This means that the fine grid evolution can
begin with three time levels of coarse grid data with which to do the
time interpolation.

We are able to obtain results from FMR simulations which compare
favourably with corresponding unigrid simulations for the following
systems: the wave equation with $1/r$ data and excision near the
origin, robust stability tests, a nonlinear gauge wave, and a
Schwarzschild black hole with excision.  The robust stability tests
indicate that numerical noise combined with inter-grid transport
operations will not lead to exponential blow ups, and that the
introduction of refinement boundaries does not produce instabilities.
Both the wave equation test and the gauge wave test show that proper
second order convergence can be obtained irrespective of waves
propagating through the interfaces between coarse and fine grids.  For
the case of the Schwarzschild black hole, although we do see an
instability at high resolutions, this instability is present in the
unigrid system as well.  At lower resolutions, we see stable behaviour
showing close agreement between FMR and unigrid results.

There are few other implementations of mesh refinement in more than
one dimension in numerical relativity with which to compare.  The
method of~\cite{Choptuik03d} is very close to ours.  However, it is an
axisymmetric code designed to evolve relativistic scalar fields.  The
code presented in~\cite{Bruegmann97} uses the ADM formalism (which is
known to be unstable) and exhibits problems at the refinement
boundaries.  In the implementation of~\cite{Joan} which uses
Paramesh~(\cite{paramesh}), all refinement levels are evolved using the
same timestep size.  This avoids the difficulties that we observe in
obtaining convergence at refinement boundaries.  However it may be
less efficient, especially where a large range of refinement levels is
employed.  Our implementation provides both the full efficiency of the
Berger--Oliger method and is a generic interface allowing the simple
implementation of other formulations and systems such as relativistic
hydrodynamics.

Finally, we note that the evolutions described in this paper were
performed by taking routines for {\em unigrid} simulations of each
physical system of interest, and after only slight modifications
to these routines, the original unigrid driver was exchanged for
the ``Carpet'' FMR driver.  The application of FMR techniques to
existing unigrid systems is thus something which, from a code
development viewpoint,  can be performed generally.
A clear advantage of this approach is that most
existing analysis tools (such as for wave extraction or apparent
horizon finding) will continue to work.
Thus we invite
other researchers to make use of our freely available code to
perform their finite-difference-based simulations in an FMR setting,
and thus achieve higher accuracies with less computing resources
than their current unigrid simulations may require.

%%%%%%%%%%%%%%%%%%%%%%%%%%%%%%%%%%%%%%%%%%%%%%%%%%%%%%%%%%%%%%%%%%%%%%%%
% Acknowledgements
%%%%%%%%%%%%%%%%%%%%%%%%%%%%%%%%%%%%%%%%%%%%%%%%%%%%%%%%%%%%%%%%%%%%%%%%
\acknowledgments We thank Bernd Br\"ugmann and Nils Dorband for their
early contributions to the work on the wave equation in FMR, Tom
Goodale, Gabrielle Allen and the Cactus team for their assistance
regarding the infrastructure development of Cactus which makes FMR
possible, and Edward Seidel for much encouragement over the past
years.  We also thank our collaborators in Wai-Mo Suen's WuGrav group,
with which we shared an early version of the code.  Finally we thank
Richard Matzner, Bernd Br\"ugmann, and Denis Pollney
for their helpful comments
regarding the preparation of this manuscript.  Work on this project
was funded by the AEI, the DFG's Collaborative Research Centre SFB
382, and by NSF grant PHY 0102204.

%%%%%%%%%%%%%%%%%%%%%%%%%%%%%%%%%%%%%%%%%%%%%%%%%%%%%%%%%%%%%%%%%%%%%%%%
% Appendixes
%%%%%%%%%%%%%%%%%%%%%%%%%%%%%%%%%%%%%%%%%%%%%%%%%%%%%%%%%%%%%%%%%%%%%%%%
\appendix
\section{Coupling coarse and fine grid time evolution}
\label{sec:bufferzones}

When certain formulations of the equations are evolved with multi-step
time integration methods, we find that the standard Berger--Oliger
approach leads to a loss of convergence at the boundaries of fine
grids. We will consider only second order methods here. We will argue
that the use of Berger--Oliger style prolongation to find boundary
conditions for refined grids at the intermediate steps causes a second
order truncation error after a fixed number of time steps. Therefore
the global error will be first order for a hyperbolic system where to
integrate to a fixed time the number of timesteps increases with the
grid resolution.

We first give an example, considering a simple one dimensional
example on an infinite domain. The wave equation
\begin{eqnarray}
\partial^2_t u & = & \partial^2_x u
\end{eqnarray}
can be reformulated into a system that is first order in time,
but still second order in space:
\begin{eqnarray}
\partial_t u & = & v \\
\partial_t v & = & \partial^2_x u \textrm{.}
\end{eqnarray}
Discretising this system on a uniform grid in space and time, we
represent the function $u$ and $v$ through
\begin{eqnarray}
u(t,x) & = & U^n_i \\
v(t,x) & = & V^n_i
\end{eqnarray}
at the collocation points $t=n \Delta t$, $x=i \Delta x$, where $n$
and $i$ number the grid points in the temporal and spatial direction,
and $\Delta t$ and $\Delta x$ are the temporal and spatial grid
spacings, respectively.  The ratio $\alpha=\Delta t/\Delta x$ is the
CFL factor. $h$ will be used to refer to terms that are order $\Delta
x$ or $\Delta t$.

For the spatial discretisation of the operator $\partial^2_x$
we will consider the second order centred differencing stencil
\begin{eqnarray}
  D^2_x F(x) & := &
  \frac{1}{(\Delta x)^2} \left( F(x-\Delta x) - 2 F(x) + F(x+\Delta x)
  \right) \textrm{.}
\end{eqnarray}

For time integration we consider one form of the ICN (iterative
Crank-Nicholson) method.  It consists first of an Euler step
\begin{eqnarray}
T_E[F_0] & := & F_0 + \Delta t\, \partial_t F_0
\end{eqnarray}
and then several (in our case, two) ICN iterations
\begin{eqnarray}
T_I[F_0, F] & := & F_0 + \Delta t\, \partial_t \frac{1}{2} \left( F_0
  + F \right) 
\end{eqnarray}
so that the overall ICN integration with $2$ iterations is given by
\begin{eqnarray}
T_{\mathrm{ICN}(2)}[F_0] & := & T_I[F_0, T_I[F_0, T_E[F_0]]] .
\end{eqnarray}
Note that there exist several slightly different variants of the ICN
method.

This discretisation is correct up to second order, which means that for
a fixed CFL factor $\alpha$, the discretisation error is $O(h^2)$.
This means also that functions $f(t,x)$, which depend on $t$ and $x$
in a polynomial manner with an order that is not higher than
quadratic, are represented exactly, i.e.\ without discretisation
error.  (Such functions $f$ can be written as
\begin{eqnarray}
\label{eqn:all-exact}
f(t,x) & := & \sum_{p=0}^2\, \sum_{q=0}^2\, C_{pq}\, t^p x^q
\end{eqnarray}
with constant coefficients $C_{pq}$.)

\subsection{Performing a single step}
\label{sec:perf-single-step}

It is illustrating to perform a time integration step by step.  We
will use for that the solution of the wave equation
\begin{eqnarray}
\label{eqn:quadratic-solution-u}
u(t,x) & = & \frac{1}{2}\,t^2 + \frac{1}{2}\,x^2 \\
\label{eqn:quadratic-solution-v}
v(t,x) & = & t
\end{eqnarray}
which will test whether the formulation is indeed capable of
representing all functions of the form (\ref{eqn:all-exact}).

Starting from the exact solution at time $t_0$,
\begin{eqnarray}
U(t_0,x) & = & \frac{1}{2}\,t_0^2 + \frac{1}{2}\,x^2 \\
V(t_0,x) & = & t_0,
\end{eqnarray}
the result of the Euler step of the ICN integration is
\begin{eqnarray}
\label{eqn:unigrid-euler-u}
U^{(0)}(t_0+\Delta t,x) & = & U(t_0,x) + \Delta t\, V(t_0,x) \\
                 & = & \frac{1}{2}\,t_0^2 + \frac{1}{2}\,x^2 + (\Delta
                 t)\, t_0 \\ 
V^{(0)}(t_0+\Delta t,x) & = & V(t_0,x) + \Delta t\, D^2_x U(t_0,x) \\
\label{eqn:unigrid-euler-v}
                 & = & t_0 + \Delta t
\end{eqnarray}
where $F^{(k)}$ denotes the result after $k$ ICN iterations.  In these
expressions, $U^{(0)}(t_0+\Delta t,x)$ differs from the true solution
$U(t_0+h,x)$ by a term $O(h^2)$.

The first ICN iteration then leads to
\begin{eqnarray}
U^{(1)}(t_0+\Delta t,x) & = & U(t_0,x) + \Delta t\, \frac{1}{2}
                                  \left( V(t_0,x) + V^{(0)}(t_0+\Delta
                                  t,x) \right) \\
                 & = & \frac{1}{2}\,t_0^2 + \frac{1}{2}\,x^2 + \Delta
                                  t\, t_0 + \frac{1}{2} (\Delta t)^2 \\
                 & = & \frac{1}{2}\,(t_0+\Delta t)^2 +
                                  \frac{1}{2}\,x^2 \\ 
V^{(1)}(t_0+\Delta t,x) & = & t_0 + \Delta t\, \frac{1}{2}
                      \left( D^2_x U(t_0,x) +
                             D^2_x U^{(0)}(t_0+\Delta t,x) \right) \\
                 & = & t_0 + \Delta t
\end{eqnarray}
which is already the correct result.  The second ICN iteration does
not change the above expressions.

\subsection{Mesh refinement}
\label{sec:mesh-refinement}

Let us now introduce mesh refinement, so that the spatial and temporal
resolution is not uniform any more.  Let the grid points be staggered
about the origin, so that the grid points with $x<0$ have a coarse
spatial grid spacing $2\Delta x$, and the grid points with $x>0$ have
a fine spatial grid spacing $\Delta x$.  Let the CFL factor $\alpha$
remain uniform, so that the temporal grid spacing is correspondingly
$2\Delta t$ for $x<0$ and $\Delta t$ for $x>0$.
We will use the Berger--Oliger time stepping scheme, which is
explained in Section~\ref{sec:time-evolution} and illustrated in
Figure~\ref{fig:Timestepping} of the main text.

The discretisation of $\partial^2_x$ needs so-called ghost
points on the fine grid, which are filled by prolongation.
Our third order polynomial
spatial prolongation operator is given by
\begin{eqnarray}
\label{eqn:prolongate-space}
P_S[F](x) & := & \left\{
\begin{array}{ll}
F(x) & \textrm{when aligned} \\
\frac{1}{16} \left[-1, +9, +9, -1\right] \\
{} \quad \cdot F(x + \left[-3,-1,+1,+3\right] \Delta x) & \textrm{otherwise}
\end{array}
                 \right.
\end{eqnarray}
and the second order polynomial temporal prolongation operator by
\begin{eqnarray}
\label{eqn:prolongate-time}
P_T[F](x) & := & \left\{
\begin{array}{ll}
F(t) & \textrm{when aligned} \\
\frac{1}{8} \left[-1, +6, +3\right] \\
{} \quad \cdot F(t + \left[-3,-1,+1\right] \Delta t) & \textrm{otherwise}
\end{array}
                 \right.
\end{eqnarray}
where $[\cdots]$ denotes a vector, and the operator $(\cdot)$ a dot
product.

Let us now consider a time evolution of the above solution
(\ref{eqn:quadratic-solution-u}--\ref{eqn:quadratic-solution-v}).
According to the Berger--Oliger time stepping scheme, the coarse grid
evolution happens as on a uniform grid.  The fine grid can now be
evolved in two different ways: (a) with a Dirichlet boundary condition
from prolongating from the coarse grid, or (b) without a boundary
condition, i.e.\ only as an IVP (as opposed to an IBVP).  In case (b),
the ``lost'' points have to be filled by prolongation after the time
step; in that case, prolongation therefore has to fill not $1$, but
$k+1$ grid points for an ICN integration with $k$ iterations.
This is illustrated in Figure~\ref{fig:bufferzones} of the main text.

\subsection{ICN integration with prolongation}
\label{sec:icn-integration-with}

Let us examine case~(a) in more detail,
in which case the fine grid boundary values
at $x<0$ are given through prolongation.  For the solution
(\ref{eqn:quadratic-solution-u}--\ref{eqn:quadratic-solution-v}), time
integration and prolongation are exact, hence it is for $x<0$
\begin{eqnarray}
  U^{(k)}(t_0+\Delta t,x) & = & \frac{1}{2}\,(t_0+\Delta t)^2 +
  \frac{1}{2}\,x^2 \\ 
  V^{(k)}(t_0+\Delta t,x) & = & t_0+\Delta t
\end{eqnarray}
for all ICN iterations $k$.  (This assumes that all ICN
iterations end up the final time.  There are also different ways of
expressing ICN.)

The Euler step then leads to, for $x>0$,
\begin{eqnarray}
  U^{(0)}(t_0+\Delta t,x) & = & \frac{1}{2}\,t_0^2 + \frac{1}{2}\,x^2 +
  \Delta t\, t_0 \\ 
  V^{(0)}(t_0+\Delta t,x) & = & t_0 + \Delta t
\end{eqnarray}
as in
(\ref{eqn:unigrid-euler-u}--\ref{eqn:unigrid-euler-v}) above.  For all
values of $x$, this can be written as
\begin{eqnarray}
  U^{(0)}(t_0+\Delta t,x) & = & \frac{1}{2}\,(t_0+\Delta t)^2 +
  \frac{1}{2}\,x^2 - \Theta(x \!>\! 0)\; \frac{1}{2}\, (\Delta x)^2 \\
  V^{(0)}(t_0+\Delta t,x) & = & t_0 + \Delta t
\end{eqnarray}
by using the $\Theta$ function, which is defined as
\begin{eqnarray}
\Theta(L) & = & \left\{
\begin{array}{ll}
0 & \textrm{when $L$ is false} \\
1 & \textrm{when $L$ is true} \textrm{.}
\end{array}
                \right.
\end{eqnarray}

The first ICN iteration leads to
\begin{eqnarray}
  U^{(1)}(t_0+\Delta t,x) & = & \frac{1}{2}\,(t_0+\Delta t)^2 +
  \frac{1}{2}\,x^2 \\ 
  V^{(1)}(t_0+\Delta t,x) & = & t_0 + \Delta t + \Theta(0 \!<\! x
  \!<\! H)\; \frac{1}{2} \alpha^2 \Delta t \textrm{.}
\end{eqnarray}
On a uniform grid, and for our solution $u$, the first ICN iteration
is already exact.  In general, the first ICN iteration should lead to
an error which is $O(h^2)$.  This is here not the case; clearly
$V^{(1)}$ has an $O(h)$ error.  However, the error is confined to a
region of length $\Delta x$, so an integral norm of the error can be
$O(h^2)$; it is not clear what this means in practice.

The result of the second ICN iteration is
\begin{eqnarray}
  U^{(2)}(t_0+\Delta t,x) & = & \frac{1}{2}\,(t_0+\Delta t)^2 +
  \frac{1}{2}\,x^2 + \Theta(0 \!<\! x \!<\! H)\, \frac{1}{4}
  \alpha^2 (\Delta t)^2 \\ 
  V^{(2)}(t_0+\Delta t,x) & = & t_0 + \Delta t \textrm{.}
\end{eqnarray}
This expression is the final result of the first fine grid time step.
It has an error that is $O(h^2)$, or $O(h^3)$ in an integral norm.
As noted above a local error that is $O(h^2)$ will lead to a global
error of $O(h)$ when considered at a fixed time. That is, the method
is only first order convergent.

A third ICN step --- that we don't perform, but it would be possible
to do so --- would result in
\begin{eqnarray}
  U^{(3)}(t_0 + \Delta t,x) & = & \frac{1}{2}\,(t_0+\Delta t)^2 +
  \frac{1}{2}\,x^2 \\ 
  V^{(3)}(t_0 + 2 \Delta t,x) & = & t_0 + \Delta t \\\nonumber
  & & {} + \left[ - 2\, \Theta(0 \!<\! x \!<\! \Delta x)
     + \Theta(\Delta x \!<\! x \!<\! 2 \Delta x) \right]\;
  \frac{1}{8} \alpha^4 \Delta t
\end{eqnarray}
which is worse than the result of the previous iteration.  The error
is again $O(h)$, but has now ``infected'' two grid points.  The error
is smaller by a factor $\alpha^2$, but the error at the two grid
points has a different sign, indicating the start of an oscillation.

It should be noted that using higher order derivatives, e.g.\ using a
five-point stencil for a fourth-order second derivative, does not
remove this error.  Similarly, using a higher order prolongation
operator does not help.  (We assume that it would be possible to adapt
the prolongation scheme directly to the time integration scheme and
arrive at a consistent discretisation in this way.  We do not think
that this is worthwhile in practice.)

The main problem seems to be caused by taking a second derivative,
which has formally an $O(1)$ error.  The usual arguments why the error
should be smaller in practice \cite{fd-consistency} do not hold near
the discontinuity that is introduced at the refinement boundary.

The result of simulating case~(a) is shown in
Figure~\ref{fig:waveeqn-bad} of the main text.  For high resolutions,
convergence seems to be only of first order.
Figure~\ref{fig:waveeqn-good} shows case~(b), which shows second order
convergence for all resolutions.

\subsection{The general case}
\label{sec:general-case}

In the general case the coarse grid evolution and the prolongation
will not provide exact data at the boundary on the fine grid. Instead
we argue heuristically as follows. The initial data $U^{(0)},V^{(0)}$
will be correct to order $h^2$,
\begin{equation}
  \label{eq:1}
  U^{n} = U(t^n, x) + h^2 e^{n} f^{n}(x) + O(h^3)
\end{equation}
where $e^{n}$ is a constant and $f^{n}$ a smooth function of $x$.
This data will be evolved forward on the coarse grid using ICN to time
$t^{n+1}$ where it will also be correct to order $h^2$,
\begin{equation}
  \label{eq:2}
  U^{n+1} = U(t^{n+1}, x) + h^2 e^{n+1} f^{n+1}(x) + O(h^3)
\end{equation}

The fine grid data will then be evolved. The first ICN step is an
Euler step to $t^{n+1/2}$ (as the fine grid timestep is one half the
coarse grid timestep). On the interior of the refined grid the result
will be first order accurate in time and second order in space,
\begin{equation}
  \label{eq:3}
  U^{n+1/2} = U(t^{n+1/2}, x) + (\Delta t e_t^{n+1/2} + (\Delta t)^2
  e_x^{n+1/2}) f^{n+1/2}(x) + O(h^3). 
\end{equation}

The boundary data for the fine grid is then found by
prolongation. Interpolation in space and time, assumed to be second
order accurate, gives
\begin{equation}
  \label{eq:4}
  U^{n+1/2}(x<0) = U(t^{n+1/2}, x) + h^2 e_p^{n+1/2}  f^{n+1/2}(x) +
  O(h^3).  
\end{equation}

Therefore at the refinement boundary there is a discontinuous jump in
the function,
\begin{equation}
  \label{eq:5}
  \left[ U^{n+1/2}\right]^{x=0+}_{x=0-} = \left(h e_t^{n+1/2} +
  (\Delta t)^2 \left( e_x^{n+1/2} - e_p^{n+1/2} \right) \right)
  f^{n+1/2}(x). 
\end{equation}
As shown in the case of simple polynomial data in
appendix~\ref{sec:icn-integration-with}, the next step in the ICN loop
will lead to a second order local error when the second derivative of
the function is taken at the discontinuity. Thus the error at a fixed
time will only be first order convergent. It is clear that this
discontinuity will also lead to first order errors with other
multi-step time integration methods such as the Runge-Kutta methods.

We have performed a symbolic calculation of two complete coarse grid
time steps with generic initial data using Maple~\cite{maple}.  We
find that the errors in $u$ and $v$ scale with $\Delta x^2$ and
$\Delta x$, respectively, both after the first and the second coarse
grid steps.  This is consistent with the calculation above as well as
our numerical results. This means that, after a fixed time $t$, the
error in $u$ is $O(\Delta x)$, so that we expect first order
convergence only.

%%%%%%%%%%%%%%%%%%%%%%%%%%%%%%%%%%%%%%%%%%%%%%%%%%%%%%%%%%%%%%%%%%%%%%%%
% Bibliography
%%%%%%%%%%%%%%%%%%%%%%%%%%%%%%%%%%%%%%%%%%%%%%%%%%%%%%%%%%%%%%%%%%%%%%%%

\bibliographystyle{apsrev}
\bibliography{method}

\end{document}